# Discovering Links for Metadata Enrichment on Computer Science Papers

*Johann Schaible, Philipp Mayr*

# Discovering Links for Metadata Enrichment on Computer Science Papers

*Johann Schaible, Philipp Mayr*

# Abstract

At the very beginning of compiling a bibliography, usually only basic information, such as title, authors and publication date of an item are known. In order to gather additional information about a specific item, one typically has to search the library catalog or use a web search engine. This look-up procedure implies a manual effort for every single item of a bibliography. In this technical report we present a proof of concept which utilizes Linked Data technology for the simple enrichment of sparse metadata sets. This is done by discovering `owl:sameAs` links between an initial set of computer science papers and resources from external data sources like DBLP, ACM and the Semantic Web Conference Corpus. In this report, we demonstrate how the link discovery tool Silk is used to detect additional information and to enrich an initial set of records in the computer science domain. The pros and cons of silk as link discovery tool are summarized in the end.

# 1   Introduction

While setting up a research bibliography, a researcher typically starts with the basic information which identifies a record, i.e. title, authors and publication date. Then a manual search has to be done for every record to complete the fragmentary record with further metadata like publishers name, DOI, issue number, page number etc. This requires a lot of manual effort. We present a proof of concept which uses semantic web technology with the objective to integrate data from different external data sources from the Linked Open Data cloud, where hundreds of various data providers share and interlink their data with other data sets. Our approach involves the discovery of links to external data sources, which contain metadata about an initial record from the same domain. The discovered links are used for the enrichment of the metadata of the initial record. We propose a generic approach which has to be done once for all records. Conceptually, we can show that the effort for gathering additional information for sparse metadata sets can be reduced.

For better understanding, the proof of concept is described along a scenario, which is stated as follows:

> *Lars, a doctoral student in computer science, has to set up a bibliography for dozen of computer science papers with the typical metadata needed for the compilation of the reference section of a paper. For every paper, only the title, the name(s) of the author(s) and the publication date are known. Instead of searching a library catalog or using a web search engine to complete the minimal bibliographic information of these records, Lars intends to discover links to external data sources published in the Linked Open Data cloud, in order to use them to retrieve information from these external data sources and thus to enrich the metadata of the records.*

The two questions arising in this scenario are:

1.   How can the external data sources be interlinked?

2.   How can the discovered links be used to enrich the metadata of the records?

The paper is structured as follows: Section 2 addresses the first question, where we describe how links can be discovered and demonstrate the usage of the link discovery tool Silk [6][1] in our scenario. Silk is used for detecting links from the initial dataset to DBLP[2], ACM[3] and the Semantic Web Conference Corpus[4]. Section 3 addresses the second question. We provide different possibilities for utilizing discovered links to enrich the metadata of a record. In section 4 the results of our scenario are presented and in section 5 the proof of concept is concluded.

---

[1] http://www4.wiwiss.fu-berlin.de/bizer/silk/
[2] http://www.informatik.uni-trier.de/~ley/db/
[3] http://dl.acm.org/
[4] http://data.semanticweb.org/

## 2 Discovering Links

The Web of Data is built upon two ideas: First, employ the RDF[5] data model to publish structured data on the Web and second, create explicit data links between resources of different data sources [2]. Links that connect data sources are represented as RDF triples, where the subject is an URI reference from one data source, while the object is a URI reference from another data source. Figure 1 shows the approach [6] for our proof of concept, where an `owl:sameAs` link, taken from the OWL[6] vocabulary, is generated semi-automatically based on the properties of each resource. This means that two resources are considered the same, when a predefined set of properties from one data source has the same values as a predefined set of properties from another data source. Predefined properties means that the user declares which properties should be considered for comparison. This can be done manually before or during the linking process. If both resources are considered the same, all data from the external data source can be used as additional information (See Figure 1). A look at our scenario helps clarifying the linking approach:

> *Lars models his records in the bibliography in simple RDF as resources with the properties "title", "author", and as "publicationDate". Now, he intends to identify external data sources comprising to complete data about the computer science papers. Assuming he found such a data source in which resources have the properties "hasTitle", "has Author", "yearPublished", "publisher", and "journal". He defines that the properties "title" and "hasTitle", "author" and "has Author", and "publicationDate" and "yearPublished" have to be compared whether they contain equal values. If the value of "title" and "hasTitle" is equal, an owl:sameAs link is set between these two properties. For the other properties to be compared this is done analogously. If all three comparisons concerning one item end in setting an owl:sameAs link within the same resource, this resource and the external resource are considered to be the same. Hereby an owl:sameAs link is set between these resources as well. As a consequence, Lars is now able to retrieve the values of the additional and "new" properties e.g. "publisher" and "journal" from the external data source.*

### 2.1 The Data Sources

A linking to an external data sources can be accomplished in two steps. First, we model our dataset and represent it in RDF. In general, this can be done by following the basic principles provided in [1]. Second, we have to identify the external data sources. This can be done in many ways using simple web search engines. However, there are repositories holding information on data sources in the Linked Open Data cloud. One of these repositories is the open-source data platform CKAN[7]. The following subsections describe the data sources selected for our scenario.

---

[5] http://www.w3.org/RDF/
[6] http://www.w3.org/TR/owl-guide/
[7] http://ckan.org/

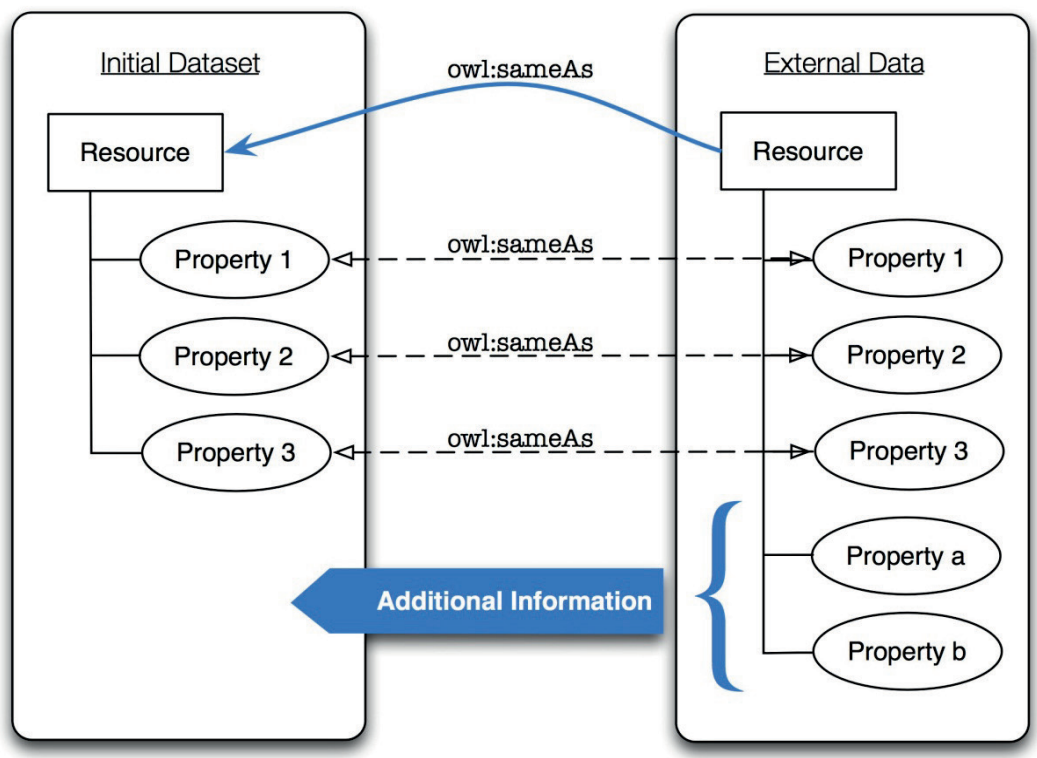

Figure 1: Interlinking Data on Instance Level

### 2.1.1   The Initial Dataset

The initial dataset was modeled according to [1], [3], and the LODE-BD recommendations[8]. It is represented in Turtle[9] syntax and utilizes the dcterms[10] as well as the FOAF[11] vocabulary. As seen in Listing 1, the title and the publication date of a paper are denoted as literals. For this, `dcterms:title` and the `dcterms:date` are used as properties. The authors on the other hand are denoted as resources of the type `foaf:Person`. Each author resource implements the properties "First Name", "Last Name" and "Full Name", which are defined as literals with `foaf:firstName`, `foaf:lastName`, and `foaf:name`. Note that all URIs in this dataset are fictional.

### 2.1.2   DBLP and ACM

By searching CKAN, we have identified the DBLP computer science bibliography and the Association for Computing Machinery (ACM) as possible external data sources including information on computer science papers. The DBLP Computer Science Bibliography provides bibliographic

---

information on computer science journals and proceedings. It indexes more than 2.1 million articles and contains many links to home pages of the authors and co-authors.

```
21  <http://lod.gesis.org/compSciencePaper/001>
22      a               <http://purl.org/linked-data/cube#DataSet>;
23      dcterms:contributor <http://lod.gesis.org/compSciencePaper/persons/paulwarren>,
24                          <http://lod.gesis.org/compSciencePaper/persons/yorksure>;
25      dcterms:creator    <http://lod.gesis.org/compSciencePaper/persons/johndavies>;
26      dcterms:date       "2011";
27      dcterms:title      "Semantic Technology and Knowledge Management".
28
29
30  <http://lod.gesis.org/compSciencePaper/persons/johndavies>
31      a               foaf:Person;
32      rdfs:label      "John Davies";
33      foaf:firstName  "John";
34      foaf:lastName   "Davies".
35
36  <http://lod.gesis.org/compSciencePaper/persons/paulwarren>
37      a               foaf:Person;
38      rdfs:label      "Paul Warren";
39      foaf:firstName  "Paul";
40      foaf:lastName   "Warren".
41
42  <http://lod.gesis.org/compSciencePaper/persons/yorksure>
43      a               foaf:Person;
44      rdfs:label      "York Sure";
45      foaf:firstName  "York";
46      foaf:lastName   "Sure".
```

Listing 1: An Instance from Lars' initial dataset

The Association for Computing Machinery (ACM) is an educational and scientific computing society, which provides publications of the ACM, along with details about the authors.

For Linked Data applications they both provide a SPARQL endpoint and a RDF dump. The RDF dump of DBLP is available as a single file with a size of about 13 gigabyte, or split up by year, resulting in files of about 200 megabyte. All data can be queried manually using the RKB explorer site[12]. Listing 2, which is a RDF/XML representation, displays an instance from this dataset. We can see that DBLP makes use of the AKT Reference[13] and the AKT Support Ontology[14]. The title is denoted as `akt:has-title`, the author name as `akt:full-name`, and the publication date of a paper is denoted using the `akts:year-of` property. Hereby the title and the publication date are modeled as literals, whereas the author is modeled as a resource with the `akt:has-author` property including the literal within `akt:full-name` property. As additional information, we can identify the journal and the web address including a DOI[15] number.

The ACM RDF dump is split by research area attributions. The single RDF files have a size of also about 200 megabyte. For users they provide a RKB explorer, which can be used to search the repository using a string based search function. All data is available on the RKB explorer site[16].

---

[12] http://dblp.rkbexplorer.com/
[13] http://www.aktors.org/ontology/portal#
[14] http://www.aktors.org/ontology/support#
[15] http://www.doi.org/
[16] http://acm.rkbexplorer.com/

Listing 3 displays an instance from this dataset, which is also represented in RDF/XML. As well as DBLP, ACM makes use of the AKT Reference and the AKT Support Ontology. We can clearly see that the data model is very similar to the one from DBLP. The only difference is the modeling of the publication date. However, this does not have an influence on the linking approach. As additional information, we detect a publication reference, which can be used to search for other scientific papers citing this instance. Furthermore, information about scientific papers this instance is citing is given also.

### 2.1.3    Semantic Web Conference Corpus

The Semantic Web Conference Corpus, also known as the "Semantic Web Dog Food Corpus", was identified as another data source, which contains information on papers that were presented, names of participants, and other things regarding the main conferences and workshops in the area of Semantic Web research. The corpus holds about 4002 papers, 9223 people, and 2726 organizations at 31 conferences and 194 workshops. For Linked Data applications the corpus provides a SPARQL endpoint and a RDF dump. The RDF dump is represented in XML and is split up by conference and workshop. Each of these snippets is again split up by the year of presentation. This results in small RDF files with a size of about 1 MB. All data is available on the FAQ webpage[17]. Listing 4 shows a RDF/XML representation of an instance from this dataset. The Semantic Web Conference Corpus uses the Dublin Core Elements vocabulary for the title and the author. For publication date they use the SWRC[18] Ontology. Hereby, the title is denoted as a literal with `dc:title` and the author is represented via the `dc:creator` property, which is a resource of the type `foaf:person`. The full name of the author is modeled using `rdfs:label`. As additional information, we can identify the category of the scientific paper as well as its abstract.

---

[17] http://data.semanticweb.org/documentation/user/faq
[18] http://swrc.ontoware.org/ontology#

```
24  <akt:Book-Section-Reference rdf:about="http://dblp.rkbexplorer.com/id/conf/birthday/DaviesWS11">
25      <owl:sameAs rdf:resource="http://dblp.l3s.de/d2r/resource/publications/conf/birthday/DaviesWS11"/>
26      <akt:has-title>Semantic Technology and Knowledge Management.</akt:has-title>
27      <akt:has-author>
28          <akt:Person rdf:about="http://dblp.rkbexplorer.com/id/people-b...53">
29              <akt:full-name>John Davies</akt:full-name>
30          </akt:Person>
31      </akt:has-author>
32      <akt:has-author>
33          <akt:Person rdf:about="http://dblp.rkbexplorer.com/id/people-8...3">
34              <akt:full-name>Paul Warren</akt:full-name>
35          </akt:Person>
36      </akt:has-author>
37      <akt:has-author>
38          <akt:Person rdf:about="http://dblp.rkbexplorer.com/id/people-d...53">
39              <akt:full-name>York Sure</akt:full-name>
40          </akt:Person>
41      </akt:has-author>
42      <akt:article-of-journal>
43          <akt:Journal rdf:about="http://dblp.rkbexplorer.com/id/journal-bcc747b8095831f06e6eb5e3c3daa10e">
44              <akt:has-title>Foundations for the Web of Information and Services</akt:has-title>
45          </akt:Journal>
46      </akt:article-of-journal>
47      <akt:cites-publication-reference rdf:resource="http://dblp.rkbexplorer.com/id/conf/birthday/2011studer"/>
48      <akt:has-web-address>http://dx.doi.org/10.1007/978-3-642-19797-0_5</akt:has-web-address>
49      <akt:has-date>
50          <akts:Calendar-Date rdf:about="http://www.aktors.org/ontology/date#2011">
51              <akts:year-of>2011</akts:year-of>
52          </akts:Calendar-Date>
53      </akt:has-date>
54  </akt:Book-Section-Reference>
```
Listing 2: An Instance from the DBLP dataset

```
38887  <akt:Article-Reference rdf:about="http://acm.rkbexplorer.com/id/1060409">
38888      <akt:has-title>Weighted primary trait analysis for computer program evaluation</akt:has-title>
38889      <akt:has-publication-reference rdf:resource="http://acm.rkbexplorer.com/id/1060409"/>
38890      <akt:has-author>
38891          <akt:Person rdf:about="http://acm.rkbexplorer.com/id/person-172993-f78b7e591a36704928f8ab6a7412a625">
38892              <akt:full-name>Lon Smith</akt:full-name>
38893          </akt:Person>
38894      </akt:has-author>
38895      <akt:has-author>
38896          <akt:Person rdf:about="http://acm.rkbexplorer.com/id/person-511046-f78b7e591a36704928f8ab6a7412a625">
38897              <akt:full-name>Jose Cordova</akt:full-name>
38898          </akt:Person>
38899      </akt:has-author>
38900      <akt:has-date>
38901          <support:Calendar-Date rdf:about="http://www.aktors.org/ontology/date#2005-06-01">
38902              <support:has-pretty-name>2005-06-01</support:has-pretty-name>
38903              <support:day-of>01</support:day-of>
38904              <support:month-of>06</support:month-of>
38905              <support:year-of>2005</support:year-of>
38906          </support:Calendar-Date>
38907      </akt:has-date>
38908      <akt:cites-publication-reference rdf:resource="http://acm.rkbexplorer.com/id/961613"/>
38909      <akt:cites-publication-reference rdf:resource="http://acm.rkbexplorer.com/id/364937"/>
38910      <akt:cites-publication-reference rdf:resource="http://acm.rkbexplorer.com/id/801059"/>
38911      <akt:cites-publication-reference rdf:resource="http://acm.rkbexplorer.com/id/187389"/>
38912      <akt:cites-publication-reference rdf:resource="http://acm.rkbexplorer.com/id/374787"/>
38913      <akt:cites-publication-reference rdf:resource="http://acm.rkbexplorer.com/id/268210"/>
38914      <akt:cites-publication-reference rdf:resource="http://acm.rkbexplorer.com/id/369354"/>
38915      <akt:cites-publication-reference rdf:resource="http://acm.rkbexplorer.com/id/980011"/>
38916      <akt:addresses-generic-area-of-interest rdf:resource="&class;K.3.2" />
38917  </akt:Article-Reference>
```
Listing 3: An Instance from the ACM dataset

```
13  <swrc:InProceedings rdf:about="http://data.semanticweb.org/conference/eswc/2011/paper/digital-libraries/1">
14      <swrc:isPartOf rdf:resource="http://data.semanticweb.org/conference/eswc/2011/proceedings"/>
15      <dc:creator rdf:resource="http://data.semanticweb.org/person/nikos-manolis"/>
16      <dc:creator rdf:resource="http://data.semanticweb.org/person/yannis-tzitzikas"/>
17      <dc:subject>Bridging keyword and advanced search</dc:subject>
18      <dc:subject>Exploratory Search and Browsing</dc:subject>
19      <dc:subject>Fuzzy Descriptions</dc:subject>
20      <dc:title>Interactive Exploration of Fuzzy RDF Knowledge Bases</dc:title>
21      <bibo:authorList rdf:resource="http://data.semanticweb.org/conference/eswc/2011/paper/digital-libraries/1/authorlist"/>
22      <swrc:abstract>There ....</swrc:abstract>
23      <swrc:month>June</swrc:month>
24      <swrc:year>2011</swrc:year>
25      <rdfs:label>Interactive Exploration of Fuzzy RDF Knowledge Bases</rdfs:label>
26      <maker rdf:resource="http://data.semanticweb.org/person/nikos-manolis"/>
27      <maker rdf:resource="http://data.semanticweb.org/person/yannis-tzitzikas"/>
28  </swrc:InProceedings>
29
30  <Person rdf:about="http://data.semanticweb.org/person/nikos-manolis">
31      <swrc:affiliation rdf:resource="http://data.semanticweb.org/organization/university-of-crete"/>
32      <swrc:affiliation rdf:resource="http://data.semanticweb.org/organization/forth-ics"/>
33      <rdfs:label>Nikos Manolis</rdfs:label>
34      <based_near rdf:resource="http://dbpedia.org/resource/Greece"/>
35      <made rdf:resource="http://data.semanticweb.org/conference/eswc/2011/paper/digital-libraries/1"/>
36      <mbox_sha1sum>7338225ba11dbbcd1a35e224d28eaa4a926e790</mbox_sha1sum>
37      <name>Nikos Manolis</name>
38  </Person>
39  <Person rdf:about="http://data.semanticweb.org/person/yannis-tzitzikas">
40      <swrc:affiliation rdf:resource="http://data.semanticweb.org/organization/university-of-crete"/>
41      <swrc:affiliation rdf:resource="http://data.semanticweb.org/organization/forth-ics"/>
42      <rdfs:label>Yannis Tzitzikas</rdfs:label>
43      <based_near rdf:resource="http://dbpedia.org/resource/Greece"/>
44      <made rdf:resource="http://data.semanticweb.org/conference/eswc/2011/paper/digital-libraries/1"/>
45      <mbox_sha1sum>e9f061dc9064b1669786bbbd730b5e1a035b8e0d</mbox_sha1sum>
46      <name>Yannis Tzitzikas</name>
47  </Person>
```

Listing 4: An Instance from the Semantic Web Conference Corpus dataset

## 2.2    Discovering Links with the Silk Framework

Silk is a link discovery framework for detecting relationships between data items within different Linked Data sources. Tools like Silk or LIMES [5] perform SPARQL queries to retrieve information. In this technical report, LIMES and other approaches for discovering links have not been taken into account. However, it should be said, that these other approaches could have worked as well to demonstrate the proof of concept.

The Silk framework[19] can be used by data publishers to set RDF links from their datasets to datasets from other data sources on the Web (see [7] as example). There are different variants of Silk for various approaches, such as using Silk on a single machine or on a cluster of machines. For the process of interlinking different data sources, Silk provides a web application called "Workbench", which can be deployed on a Tomcat server or ran from command line as a Jetty server. Discovering links with Silk is a procedure split up into a few steps.

First, all data sources which should be linked, have to be specified by the user. Hereby, the user can specify whether he wants to use a SPARQL Endpoint or a RDF dump of a dataset. Supported formats for RDF dumps are "RDF/XML", "N-Triple", "Turtle", "TTL", and "N3". The user also has to specify the prefixes of the dataset. Silk needs these prefixes to perform queries on the dataset. The final input is the linking task. Here the user defines the source dataset and the target dataset, as well as the linking task. The source and target datasets can be chosen from all the specified ones, whereas the linking task has to be an object property from an RDF, OWL or SKOS vocabulary linking individuals to individuals.

---

[19] http://wifo5-03.informatik.uni-mannheim.de/bizer/silk/

Once the data source and the linking tasks are specified, the user can open such a task to set rules for the comparison of properties. First, the user defines the properties, e.g. compare `dcterms:title` in data source *a* with `rdfs:label` in data source *b*. He can choose the properties from the property path Silk offers, as it is seen in Figure 2 on left side, where the dataset is taken from our scenario. In addition, Silk provides transformations of the values of the properties, e.g. the transformation of the value to lower case, and offers a set of comparison techniques, like using Levenshtein distance between both values, and provides aggregation techniques, like the aggregation of the minimum numerical value. These operations are displayed on the right side of Figure 2. E.g.the user chooses the "Equality" comparator, then Silk compares all values from `dcterms:title` with all values from `rdfs:label` whether they are the equal or not. Figure 3 displays the more complex definition of such rules in our scenario.

If a value is equal, Silk sets the user-defined object property between these two resources, e.g. `owl:sameAs`. Silk can display these discovered links or even store them into a file. It will contain triples like the one shown in Listing 5.

All these specifications make a link discovery very customizable and precise[20]. Let us look at the scenario once again:

> *Lars uses Silk to discover links to DBPL, ACM, and the Semantic Web Conference Corpus. He first loads all datasets into Silk as RDF dumps and specifies their namespaces. Having done this, he defines 3 `owl:sameAs` linking tasks. Hereby each task links individuals from his initial dataset to individuals from 1) DBLP, 2) ACM, and 3) to the Conference Corpus. Starting with the first linking task, he defines `dcterms:title` from his internal dataset and `akt:has-title` from the DBLP dataset as properties for comparison. He defines the Levenshtein Distance as a comparator and sets the threshold to "3", which allows slight differences in the individual's string value. He performs this process again with the properties `foaf:name` and `akt:full-name`. For the publication date on the other hand, he sets `dcterms:date` to be compared with `akts:year` utilizing the "Date Equality" comparator. As an aggregation, he uses "Minimum". This linking task is displayed in Figure 3. After setting up this linkage rule, Silk exports all discovered links between his dataset and DBLP into an output file. Listing 5 shows an instance from this output file. Lars repeats this entire procedure for the second and third linking task. The second output file contains `owl:sameAs` links to individuals from ACM, and the third one `owl:sameAs` links to the Semantic Web Conference Corpus.*

---

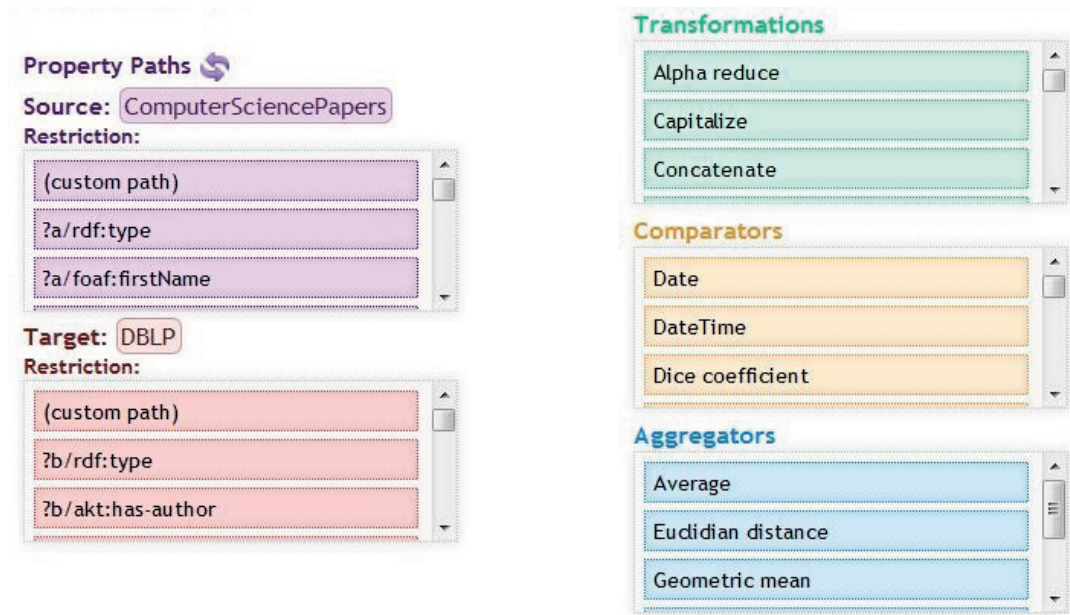

Figure 2: Silk Workbench – The Property Paths and Transformations,
Comparators, and Aggregators

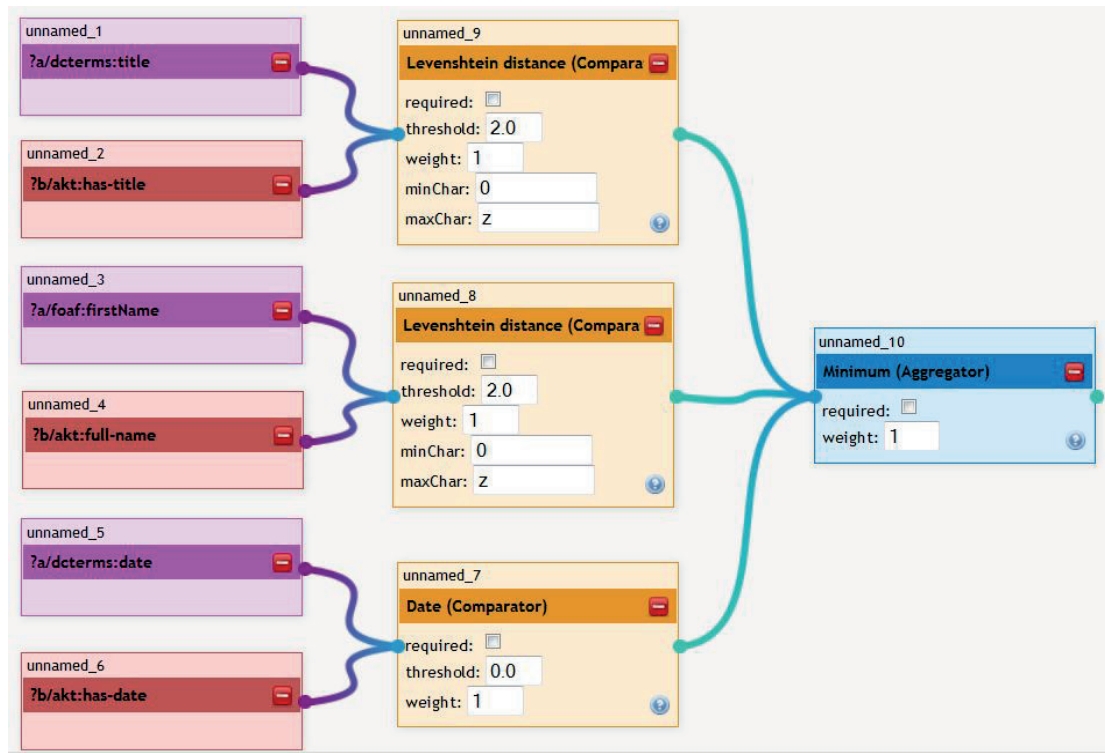

Figure 3: Silk Workbench - defining linkage rules

```
<http://lars.org/Paper/001>
    <http://www.w3.org/2002/07/owl#sameAs>
        <http://dblp.rkbexplorer.com/id/conf/birthday/DaviesWS11> .
```

Listing 5: Excerpt from a Silk output file

## 3   Enriching Metadata Using the Discovered Links

To enrich metadata means that we want to increase the "value" or improve the "quality" of an initial metadata set by adding further information. The discovered links help to retrieve further information from external data sources for metadata enrichment. Hereby, we have to distinguish between 2 types of enrichment:

1. Addition of the discovered links to the initial dataset, to make hyper references to external data sources.

2. Utilization of links to perform queries on external data sources, in order to add supplementary metadata properties to the initial dataset.

To add links to a dataset is the least possible enrichment. To follow these links would lead to all further information provided by other data publishers, without any search. The main advantage is the minimal effort needed to add the discovered links to the initial dataset. This can even be done manually without much effort. For an automatic approach, one needs to perform a merge routine, e.g. using XSLT[21]. Another advantage of integrating the links is the minimized administration effort of the resulting dataset. If an external data provider changes its metadata, it will typically not have an impact on the URI of the resource. This means, that the discovered link will lead to the most updated metadata. However, this is also the main disadvantage. By integrating only the links, one has to rely on the external data provider. Some data providers may not only change the metadata but also the domain, URIs, or property names although it is not supposed to be done. According to [4], this does happen though from time to time. This can result in syntactically and semantically wrong links. If they are syntactically broken, a browser cannot dereference them. If they are semantically incorrect, following these links will lead to a wrong resource, which will result in wrong metadata information. Dereferencing of the link also includes the presentation of the metadata information. This should be done in a human readable form. Some provide simple HTML web pages, some provide a RKB explorer, and in some cases following the link leads to a XML representation.

---

[21] http://www.w3.org/TR/xslt

```
10  <http://lars.org/Paper/001>
11  a                  <http://purl.org/linked-data/cube#DataSet>;
12  dcterms:contributor <http://lars.org/persons/paulwarren>,
13                      <http://lars.org/persons/yorksure>;
14  dcterms:creator    <http://lars.org/persons/johndavies>;
15  dcterms:date       "2011";
16  dcterms:title      "Semantic Technology and Knowledge Management";
17
18  owl:sameAs         <http://dblp.rkbexplorer.com/id/conf/birthday/DaviesWS11>.
19
20  <http://lars.org/persons/johndavies>
21      a                  foaf:Person;
22      foaf:firstName  "John";
23      foaf:lastName   "Davies";
24      foaf:name       "John Davies".
25  <http://lars.org/persons/paulwarren>
26      a                  foaf:Person;
27      foaf:firstName  "Paul";
28      foaf:lastName   "Warren";
29      foaf:name       "Paul Warren".
30  <http://lars.org/persons/yorksure>
31      a                  foaf:Person;
32      foaf:firstName  "York";
33      foaf:lastName   "Sure";
34      foaf:name       "York Sure".
```

Listing 6: the initial record is enriched with a owl:SameAs link

To use the links to perform a query on the external data sources and to add their metadata to the initial dataset is a more sophisticated type of metadata enrichment. The result is that all information is stored in one single place. In contrary to the first type of enrichment, a visit to the external data provider's web page is not required in order to get additional information. This also implies that one has not to rely on the external data provider. The disadvantage of this type of enrichment is data redundancy. An initial effort is needed to insert all the additional metadata, and even more effort is needed to update and merge, if the external data providers change their data. To integrate metadata from external data source, either a SPARQL query is needed, or a merging routine with XSLT or other transformation languages. This effort has to be done uniquely for every external data source, as most of them use a different schema. A SPARQL query is needed, if an external data source uses a SPARQL endpoint, whereas a merging routine might be slightly easier, if one used a RDF dump. It is most likely that an expert is needed in both cases, since such SPARQL queries or merging techniques can be very complicated. This results in a much higher effort than simply add links. Let us look at the scenario for the last time:

```
10  <http://lars.org/Paper/001>
11  a                  <http://purl.org/linked-data/cube#DataSet>;
12  dcterms:contributor <http://lars.org/persons/paulwarren>,
13                       <http://lars.org/persons/yorksure>;
14  dcterms:creator     <http://lars.org/persons/johndavies>;
15  dcterms:date        "2011";
16  dcterms:title       "Semantic Technology and Knowledge Management";
17
18  owl:sameAs          <http://dblp.rkbexplorer.com/id/conf/birthday/DaviesWS11>;
19
20  akt:article-of-journal  "Foundations for the Web of Information and Services";
21  akt:has-web-address     <http://dx.doi.org/10.1007/978-3-642-19797-0_5>.
22
23  <http://lars.org/persons/johndavies>
24      a               foaf:Person;
25      foaf:firstName  "John";
26      foaf:lastName   "Davies";
27      foaf:name       "John Davies".
28  <http://lars.org/persons/paulwarren>
29      a               foaf:Person;
30      foaf:firstName  "Paul";
31      foaf:lastName   "Warren";
32      foaf:name       "Paul Warren".
33  <http://lars.org/persons/yorksure>
34      a               foaf:Person;
35      foaf:firstName  "York";
36      foaf:lastName   "Sure";
37      foaf:name       "York Sure".
```

Listing 7: the initial record is enriched with data from external data sources

*Lars adds the discovered links to his dataset manually. Listing 6 shows how the initial dataset looks like, if he inserts only the links to the external data source (row 18). In this case, Silk was able to find just one link, i.e. to the DBLP dataset. This means, that the ACM and Semantic Web Conference Corpus did not have metadata on this individual. On the other hand, Listing 7 illustrates the result, where Lars applied the second type of enrichment. It shows, that all metadata from DBLP about this item is inserted into the initial dataset (rows 20-21).*

# 4 Lessons Learned Using Silk

## 4.1 Usability

To use Silk as a link discovery software has lots of positive aspects but also some shortcomings. In general the Silk Workbench is very well structured and intuitively to use. Some features are very helpful, e.g. the auto-complete function when a user types in the required prefixes. Silk detects the prefix, and suggest the vocabulary the user might intend to use. Other standard features like this are missing though, such as a browse function for defining a RDF dump as data source. The user has to type in the exact name of the file and its type. Furthermore, loading big RDF dumps (>50MB) takes some time for Silk, but it does not show any progress bar, which makes it hard for the user to see, if Silk loads the RDF dump or not. Once all data sources are loaded, Silk does present them within a tree view. This makes it very easy for the user to locate a specific data source, if a lot of sources are integrated. The linking tasks are also displayed within the same tree view with a different icon that makes it easy to distinguish it from data sources. Silk makes it easy for the user to define a linking task. The user can choose the source and target dataset, by using a drop-down box. Also he can specify from which point of the RDF graph the data shall be queried, but this was not considered in our scenario.

The definition of linkage rules seems easy at the beginning, but Silk has its own syntax the user has to get used to. All properties used in the source and target dataset are shown in a property path window. This is helpful for the user to identify the desired properties he wants to mark for comparison. If not all properties were able to load, Silk provides a "custom path", where the user can manually specify a property. Although for Silk this helps the user, it is still unclear, how the properties are nested inside one another, as Silk provides only a list. This might confuse the user, if a property is used more than once within one dataset, e.g. describing a label using `rdfs:label`. This might result in a wrong linkage rule.

The drag-and-drop functionality is very user-friendly and the way to connect two properties with a comparator is straightforward. Silk also provides an on-the-fly checking mechanism, which validates your linkage rules. Only if it validates correctly, the user can continue to generate links. This helps the user sufficiently, especially since every error, which does not pass the validation, has its own error message. Such a message displays exactly how the user can correct the error.

When a user clicks on start to generate links, Silk displays a progress bar which indicates whether the comparing process is still running or not. As a result, Silk either finds links, or it displays an empty screen. If it finds links, it displays them and writes them into a specified output file. The user can unfold each link, and check if it is a correct result or not.

## 4.2 Results

In the process of our work, we have designed some test cases, in order to check the correctness as well as the completeness of the results using Silk for link discovery. First, we set each data set as source and target dataset, meaning, that each dataset was compared with itself. We have specified the properties for comparison and started to generate links. As a result, Silk compared all values of the specified properties correctly and generated all links, which could be found. The next test case involved two datasets with a different schema. This test case was the prototype for our scenario, and it included 3 computer science papers modeled differently.

Silk has also a linkage learning function, by which the user initiates an automatic comparison of several properties with only one click; Silk then suggests the user which properties could contain the same semantic content. This function has not worked though, and thereby was not considered as part of this work.

In general we can state, that Silk is a relatively easily to use link discovery tool, which in most cases finds the matches it is supposed to find. The user faces only a few challenges, including knowing the schema of all datasets, meaning the hierarchical structure and at least a few example instances, as well as the Silk linking language, in order to discover links to external data sources.

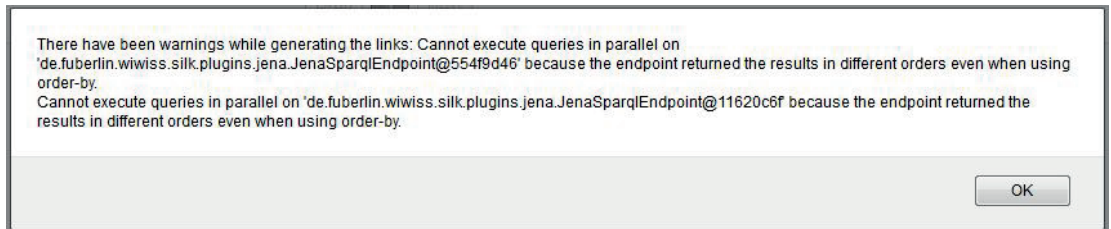

Figure 4: Silk Workbench – Error message while generating links

# 5    Conclusion

In this proof of concept, we showed two ways to enrich bibliographical datasets by discovering links to external data sources within the Linked Open Data cloud. We outlined a scenario in which computer science papers are enriched with links to the DBLP, ACM and the Semantic Web Conference Corpus. The use of computer science papers is a possible example to demonstrate the proof of concept. However, it is unclear which results might come out for data from other domains, e.g. social sciences, as it is based on the Linked Open Data cloud and the data provided there.

The modeling of the initial dataset should be done according to best practices provided by [1] in order to maximize the possibility of finding links to external data sources. For example, it cannot be disambiguated between two equal property names with different content within the same hierarchical level, i.e. do not use `rdfs:label` for all labels within the same hierarchical level.

In order to use external datasets appropriately, each dataset has to be analyzed on both, the schema and the instance level. The schema of a dataset indicates what kind of metadata, e.g. a title or a date, is modeled. However, it does not indicate whether the metadata is modeled as a literal or a resource. As Silk needs a literal for a comparison, it is necessary to examine a few instances of the dataset. Only this way the user can identify the properties which he or she intends to compare.

SPARQL is a powerful query language, which can extract information lying deep in the data. Although this is very useful in general, for our use case it is too sophisticated, as most of its features are not needed, but the effort to design queries is still enormous. Thus, for a simple enrichment it is recommended to use easier approaches like XSLT or even to alter the dataset manually. Overall, we can say that a data provider, who intends to enrich his or her initial dataset, must have some knowhow in RDF, XML, SPARQL, a link discovery tool, and Linked Data in general.

As future work we intend to perform metadata enrichment on a larger dataset of computer science papers, in order to test the performance, completeness, and the correctness of this approach in more detail as well as examine if similar results can be achieved with scientific publications from other domains. In addition, we plan to explore which best practices for modeling a Linked Open Data dataset are (most) relevant for an easy link discovery.